\documentclass[twocolumn,showpacs,showkeys,preprintnumbers,amsmath,amssymb]{revtex4}
\usepackage[lofdepth,lotdepth,caption=false]{subfig}
\usepackage{graphicx}% Include figure files
\usepackage{dcolumn}% Align table columns on decimal point
\usepackage{bm}% bold math

\usepackage{booktabs}
\usepackage{amsmath}
\usepackage{epstopdf}

\def\beq{\begin{equation}}
\def\eeq{\end{equation}}

\begin{document}

\preprint{Submitted to Chinese Physics C}

\title{Analysis of the microbunching instability in a mid-energy electron linac}% Force line breaks with \\

\author{Dazhang Huang}
\email{huangdazhang@sinap.ac.cn}
\author{Qiang Gu}
\author{Zhen Wang}
\author{Meng Zhang}
\affiliation{%
Shanghai Institute of Applied Physics\\ Chinese Academy of Sciences\\ Shanghai, 201800, China P.R.\\
}%

\author{King Yuen Ng}
\affiliation{
Fermi National Accelerator Laboratory\\ Batavia, Illinois 60510, USA\\
}

\date{\today}% It is always \today, today,

\begin{abstract}

Microbunching instability usually exists in the linear accelerator (linac) of a free electron laser (FEL) facility. If it is not controlled effectively, the beam quality will be damaged seriously and the machine will not operate properly. In the electron linac of a soft X-Ray FEL device, because the electron energy is not very high, the problem can become even more serious. As a typical example, the microbunching instability in the linac of the proposed Shanghai Soft X-ray Free Electron Laser facility (SXFEL) is investigated in detail by means of both analytical formulae and simulation tools. In the study, a new mechanism of introducing random noise into the beam current profile as the beam passes through a chicane-type bunch compressor is proposed.  The higher-order modes that appear in the simulations suggest that further improvement of the current theoretical model of the instability is needed. 
\end{abstract}

%Use showkeys class option if keyword
                              %display desired\vskip-0.2in
\maketitle

\section{\label{sec:level1}Introduction}

The microbunching instability that occurs in the linac of a free-electron Laser (FEL) facility has always been a problem that degrades the quality of electron beams. The instability is driven by various effects, such as the longitudinal space charge (LSC)~\cite{Saldin}, coherent synchrotron radiation (CSR)~\cite{Heifets,ZHuang1}, and linac wakefields. As the beam passes through a bunch compressor, for example, a magnetic chicane, the energy modulation introduced by those effects is transformed into density modulation and thus the instability develops. In a FEL facility, there is usually more than one bunch compressor and the overall growth of the instability is the product of all the gains in each compressor. As a result, the final gain of the instability can become significant. On the other hand, the FEL process has a high demand for electron beam quality in terms of peak current, emittance, energy spread, etc. Therefore without effective control, the microbunching instability can damage the beam quality so seriously that the whole FEL facility fails.

The Shanghai Soft X-Ray FEL Facility (SXFEL) project has been proposed and the construction will start soon. It is a cascading high-gain harmonic generation (HGHG) FEL facility operating at the wavelength of 9~nm in the soft-X-ray region with 840 MeV electron beam energy at the linac exit. In this article, as a typical example, the study of the microbunching instability based on the design parameters of the SXFEL linac is carried out in order to gain better understanding of the instability process and to analyze the limitations of analytic formulae and numerical simulations. The study shows that the initial noise level of the beam current has a big impact on the final behavior of the instability, which makes the analysis of the initial current noise imperative. On the other hand, we realize that higher-order harmonics of the beam current can be excited in a bunch compressor, which may modify the linear-theory modeling of the gain. Moreover, the noise in one transverse plane can also be transferred into the longitudinal direction due to the transverse-longitudinal coupling in each dipole magnets within a magnetic bunch compressor.  This transfer reduces the smoothness of the current and its effect will be considered. Also because of this, extra noise can be introduced into the beam current when the beam passes through a chicane-type laser heater. The structure of this paper is as follows: In Sec.~II, we introduce the basic structure of the SXFEL linac and the fundamental theory of microbunching instability. Comparisons of the gain curves with and without the initial current noise are provided based on the present design parameters and the results are analyzed. In Sec.~III, the higher-order modes of the instability exhibited in the study are presented, In Sec.~IV, a new mechanism of introducing random noise into the beam current is proposed and discussed. Summaries and concluding remarks are given in Sec.~V.       

\section{Microbunching instability in SXFEL}

The basic mechanism of the microbunching instability driven by various effects has been studied and their corresponding theories have been derived in Refs.~\onlinecite{Saldin,ZHuang,Heifets}. The development of the instability is similar to the amplification process in a klystron amplifier. The initial density modulation and/or white noise are transformed into energy modulation by various kinds of impedance (e.g., LSC, CSR, linac wakefields) during beam transportation. The energy modulation accumulates and is fed-back into the beam after passing through a dispersive section such as a bunch compressor. This accumulation results in much stronger amplified density modulation. Additionally, the CSR effect in the dispersive section forms a positive feedback which in turn enhances the instability. More dispersive sections in a linac result in stronger instability growth. %The whole process is illustrated in Figure~\ref{amplification}.

\begin{figure}[htb]
   \centering
   \includegraphics*[width=80mm]{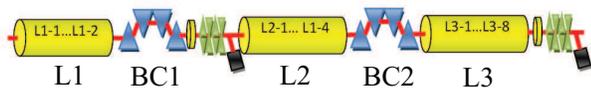}
\vskip-0.05in
   \caption{(Color) Layout of the SXFEL linac.}
   \label{linac}
\end{figure}

\begin{figure}[htb]
   \centering
   \includegraphics*[width=70mm]{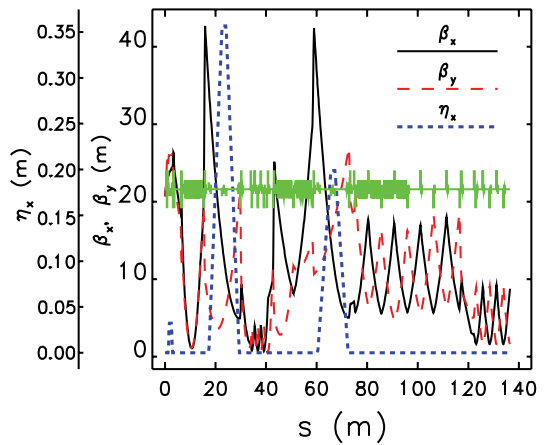}
\vskip-0.05in
   \caption{(Color) Magnet layout and Twiss parameters of the SXFEL linac.}
   \label{magnet}
\end{figure}

In the linac of the SXFEL, both S-band and C-band rf structures are used to accelerate the electron beam. One X-band structure is implemented to suppress the second-order nonlinear components in the longitudinal phase space to avoid undesired growth of transverse emittance and slice energy spread. Two magnetic chicane-type bunch compressors (BC1 and BC2) are used to compress the beam to arrive at the required peak current. The layout of the SXFEL linac is shown in Figure~\ref{linac}, and the magnet layout and some of the Twiss parameters of the SXFEL linac are shown in Figure~\ref{magnet}.  Note that S-band accelerating structures are used in sections L1 and L2, whereas a C-band accelerating structure is used in section L3. The following discussions are based on the design parameters shown in Table~\ref{parameters}, which is taken from the SXFEL feasibility study report. Moreover, since the length scale in which the structural impedance is effective is much longer than that of microbunching wavelength~\cite{feasibility,Venturini1}, we may neglect the effects from the linac wakefields in the following discussions without compromising accuracy. 

\begin{table}
\caption{\label{parameters}Main beam parameters used in the microbunching-instability study for the SXFEL (obtained by {\footnotesize ELEGANT}~\cite{elegant}, with laser heater turned off in simulation).}
\begin{ruledtabular}
\begin{tabular}{lllll}
 Parameter & Value \\
\hline
bunch charge (nC) & 0.5 \\ 
beam energy out of injector (MeV) & 130 \\
bunch length (FWHM) at the exit of injector (ps) & 8 \\
peak current before BC1 (A) & 60\\
beam radius (rms) before BC1 (mm) & 0.40 \\
beam energy before BC1 (MeV) & 208.4 \\
local (slice) rms energy spread right before BC1 (keV)\!& 6.5 \\
linac length up to BC1 (m)& 17.3 \\
$R_{56}$ of BC1 (mm) & 48 \\
beam energy before BC2 (MeV) & 422.0 \\
local (slice) rms energy spread right before BC2 (keV)\! & 43.4 \\
linac length up to BC2 (m)& 60.3 \\
$R_{56}$ of BC2 (mm) & 15 \\
linac length after BC2 (m)& 63.7 \\
compression ratio (BC1$\times {\rm BC2}$) & $5\times 2$ \\
beam radius (rms) before BC2 (mm) & 0.26 \\
\end{tabular}
\end{ruledtabular}
\end{table}

\begin{figure}[htb]
\centering
\subfloat[The longitudinal current profile of the input beam.]{
   \includegraphics[width=0.70\linewidth]{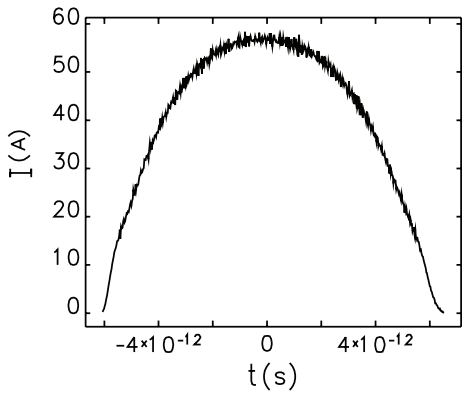}
   \label{subfig:fig1}
}
%\subfloat[short for lof][long subfig2 caption]{
%   \includegraphics[width=0.45\linewidth]{figure2.png}
%   \label{subfig:fig2}
%}

\subfloat[The longitudinal phase space distribution of the input beam.  Note that the unit of momentum is electron mass divided by the speed of light.]{
   \includegraphics[width=0.70\linewidth]{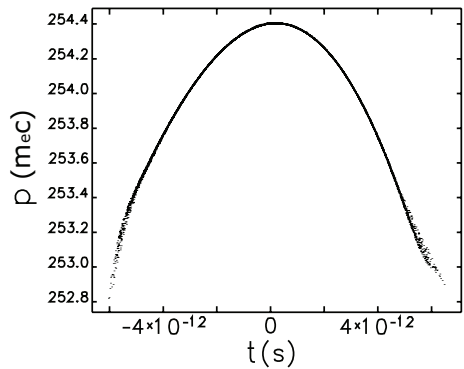}
   \label{subfig:fig3}
}
%\subfloat[short for lof][long subfig4 caption]{
%   \includegraphics[width=0.45\linewidth]{figure4.png}
%   \label{subfig:fig4} 
%}
\caption[short for lof]{The input beam current profile (a) and the longitudinal phase space (b).}
\label{inputbeam}
\end{figure}

The microbunching instability for the case of linear compression has been discussed by Saldin et~al.~\cite{Saldin} phenomenologically by comparing the energy distributions before and after the compression. Consider a density modulation at wavenumber $k$.  Without higher harmonics of beam current taken into account,  the gain of the instability driven by the wake fields upstream of the compressor reads~\cite{Saldin} 
\beq
G=Ck|R_{56}|\frac{I_0}{\gamma I_A}\frac{|Z_{\rm tot}(k)|}{Z_0}\exp\Bigg(-\frac{1}{2}C^2k^2R^2_{56}\frac{\sigma^2_\gamma}{\gamma^2}\Bigg).
\label{saldin}
\eeq
Here, $\gamma$ is the nominal relativistic factor of the electron beam with rms local energy spread $\sigma_\gamma$ in front of the bunch compressor, $C=1/(1+hR_{56})$ is the compression ratio, $h$ is the linear energy chirp, $R_{56}$ is the 5-6 element of the transport matrix, $I_0$ is the initial peak current of the beam, $Z_0=377 \ \Omega$ is the free-space impedance, $Z_{\rm tot}$ is the overall impedance upstream of compressor including those of the LSC, linac wake, etc., and $I_A=17$~kA is the Alfven current. The longitudinal space-charge impedance per unit length in free space takes the form~\cite{ZHuang,Venturini}:
\begin{align}
Z_{\rm LSC}(k)&=\frac{iZ_0}{\pi kr^2_b}\Bigg[1-\frac{kr_b}{\gamma}K_1\Bigg(\frac{kr_b}{\gamma}\Bigg)\Bigg]
\nonumber \\
&\approx\left\{\begin{array}{lll}
\frac{iZ_0}{\pi kr^2_b} & \frac{kr_b}{\gamma}\gg1,\\
~&~ \\
\frac{iZ_0k}{4\pi\gamma^2}(1+2{\rm ln}\frac{\gamma}{r_bk}) & \frac{kr_b}{\gamma}\ll1,
\end{array}\right.
\label{LSCimp}
\end{align}
where $r_b$ is the radius of the beam, $K_1$ is the first-order modified Bessel function of the second kind. Based on Eqs.~(\ref{saldin}) and (\ref{LSCimp}), using the parameters output by the {\footnotesize ELEGANT} simulation starting from a beam with $\sim\pm1\%$ noise fluctuation in current and 1 -- 2 keV uncorrelated (slice) energy spread (Figure~\ref{inputbeam}),  the gains of microbunching instability in the region around the peak current induced by LSC impedance at the exits of BC1 and BC2 (first and second bunch compressors) are computed by the analytic formula. The beam parameters at the exit of the SXFEL injector are prepared by {\footnotesize PARMELA}~\cite{feasibility,parmela} simulation employing one million macro-particles. The LSC impedances are computed separately in drift space and the accelerating section.

The reason why we purposely choose to use a noisy input instead of a smooth one is because the real beam is usually not ideally smooth, it always includes the ripples introduced by the random noise fluctuation, the quantum process of field emission, the laser power jitter, etc. On the other hand, the linear theory is applied in the computation because: I. in general, the initial density modulation is very small, and II. the gain in the first compressor BC1 is small as well. 

Besides the LSC-induced microbunching instability, when a charged particle beam passes through a bunch compressor, coherent synchrotron radiation (CSR) can be emitted at wavelengths much shorter than the bunch itself.  The density of bunch particles is modulated at those wavelengths. As we have already discussed, since the CSR effect in a bunch compressor introduces positive-feedback to the microbunching instability, the gain of the instability therefore rises rapidly.  

The gain of CSR-driven microbunching instability in a bunch compressor has been derived analytically~\cite{ZHuang1} in terms of beam energy, current, emittance, energy spread and chirp, as well as initial lattice and chicane parameters. Assuming a beam uniform in the $z$-direction and Gaussian in transverse and in energy distributions, the CSR-driven microbunching instability growth follows the expression~\cite{ZHuang1}:  
\begin{align}
G_f&\approx\bigg |\exp\bigg [-\frac{{\bar{\sigma}}^2_\delta}{2(1+hR_{56})^2}\bigg ]+A{\bar{I}}_f\bigg [\bigg (F_0({\bar{\sigma}}_x) \nonumber \\
&+\frac{1-e^{-{\bar{\sigma}}^2_x}}{2{\bar{\sigma}}^2_x}\bigg )\exp{\bigg (-\frac{{\bar{\sigma}^2_\delta}}{2(1+hR_{56})^2}\bigg )} \nonumber \\
&+F_1(hR_{56},{\bar{\sigma}}_x,\alpha_0,\phi,{\bar{\sigma}}_\delta)\bigg ] \nonumber \\
&+A^2{\bar{I}}^2_fF_0({\bar{\sigma}_x})F_2(hR_{56},{\bar{\sigma}}_x,\alpha_0,\phi,{\bar{\sigma}}_\delta)\bigg |.
\label{CSRgain}
\end{align}

As described in Ref.~\onlinecite{ZHuang1}, the first term on right side of Eq.~(\ref{CSRgain}) represents the loss of microbunching in the limit of vanishing current, the second term (linear in current) provides a one-stage amplification at low current (low gain), and the last term (quadratic in current) corresponds to the two-stage amplification at high current (high gain). The functions of $F_0$, $F_1$, and $F_2$ are defined in Ref.~\onlinecite{ZHuang}, and $\bar{\sigma}_\delta$ and $\bar{\sigma}_x$ are related to, respectively, the local (slice) energy spread and rms transverse size of the beam. 

In our calculation for the SXFEL linac, since the high-gain term in Eq.~(\ref{CSRgain}) is much smaller than the low-gain term (possibly due to the rapid increase of slice energy spread in the bunch compressor), we just ignore the contribution of the two-stage amplification.

\begin{figure}[htb]
   \centering
   \includegraphics*[width=60mm]{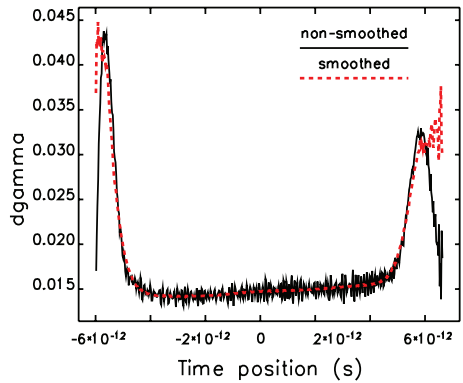}
\vskip-0.10in
   \caption{(Color) The slice energy spreads of the smoothed (red dot) and the non-smoothed beam (black) before BC1.}
   \label{L1dgamma}
\end{figure}

\begin{figure}[htb]
   \centering
   \includegraphics*[width=60mm]{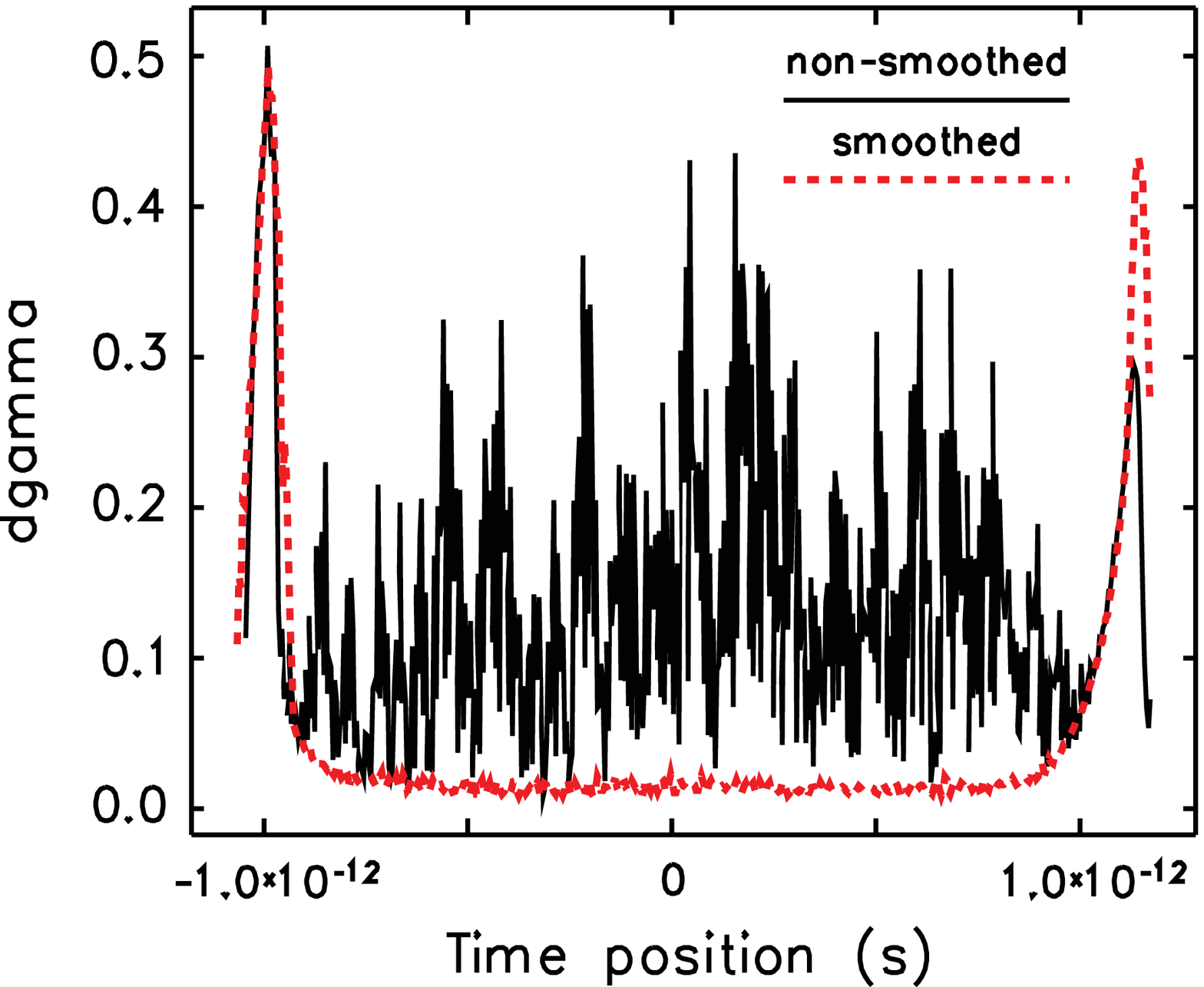}
\vskip-0.10in
   \caption{(Color) The slice energy spreads of the smoothed (red dots) and the non-smoothed beam (black) before BC2.}
   \label{L2dgamma}
\end{figure}

\begin{figure}[htb]
   \centering
   \includegraphics*[width=60mm]{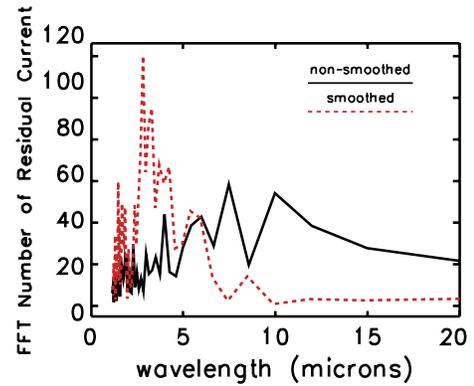}
\vskip-0.10in
   \caption{(Color) The current spectra of the smoothed (red dots) and non-smoothed beam (black) at the exit of the SXFEL linac.  The wavelength is reduced by the factor of 10 from the initial value because of compression.}
   \label{finalcurrentFFT}
\end{figure}

As we mentioned in the previous section, noise is always introduced inevitably when electrons are emitted from the cathode. Therefore, in order to have a clearer picture about the effect of the initial noise on the final gain, the gain with an initially smoothed electron beam is computed as well. Figure~\ref{inputbeam}\subref{subfig:fig1} shows the current profile of the input beam with $\sim1\%$ noise fluctuation in rms, and figure~\ref{inputbeam}\subref{subfig:fig3} is the longitudinal phase space of the beam.  Our study shows that the uncorrelated (slice) energy spread before the second bunch compressor (BC2) becomes larger for an initially noisy (non-smoothed) beam input.  As a result, its final gain is smaller than the one computed with an initially smoothed beam. Figures~\ref{L1dgamma} and \ref{L2dgamma} show, respectively, the uncorrelated (slice) energy spread of the smoothed and the non-smoothed beam before BC1 and BC2 by {\footnotesize ELEGANT}. All the parameters in the analytical calculation are obtained from {\footnotesize ELEGANT}. The term ``smoothed beam'' means the whole six-dimensional beam distribution is smoothed by removing the noise fluctuation, which is done by the sdds command smoothDist6s~\cite{sddstoolkit}. With this command, the noise fluctuation of the beam current profile in both the longitudinal and the transverse directions, and that of the uncorrelated (slice) energy spread are all suppressed to almost zero. As we have already known, due to the smaller slice energy spread, the peak gain of the smoothed beam is much higher than that of the noisy beam, and the wavelength where the peak resides is also much shorter. However, although the total gain is small for the noisy beam, the microbunching instability can still be a problem because of the significant initial noise. As a demonstration, Figure~\ref{finalcurrentFFT} shows that the final current fluctuation of the non-smoothed beam is comparable to that of the smoothed one.  Thus it is also consistent with the analytical results (Figure~\ref{gaincurve}) in terms of the wavelength where the peak resides. In summary, careful studies are needed to decide how much noise should be included in the computation.

\begin{figure}[htb]
   \centering
   \includegraphics*[width=80mm,angle=0]{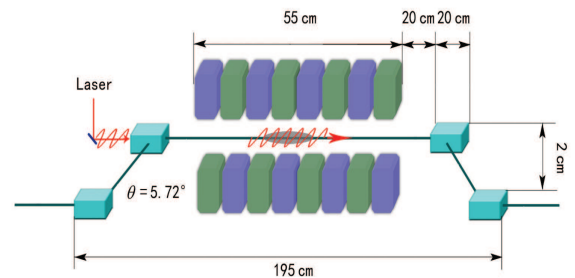}
\vskip-0.10in
   \caption{The layout of the SXFEL laser heater at 130 MeV.}
   \label{laserheater}
\end{figure} 

\begin{figure}[htb]
\centering
\subfloat[The gain curves of a noisy beam at various laser power.]{
   \includegraphics[width=0.70\linewidth]{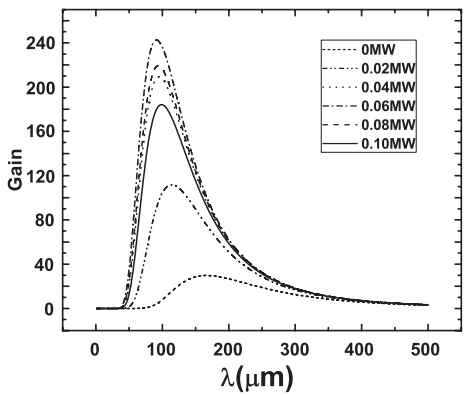}
   \label{subfig:fig1-8}
}
%\subfloat[short for lof][long subfig2 caption]{
%   \includegraphics[width=0.45\linewidth]{figure2.png}
%   \label{subfig:fig2}
%}

\subfloat[The gain curves of a smoothed beam at various laser power.]{
   \includegraphics[width=0.70\linewidth]{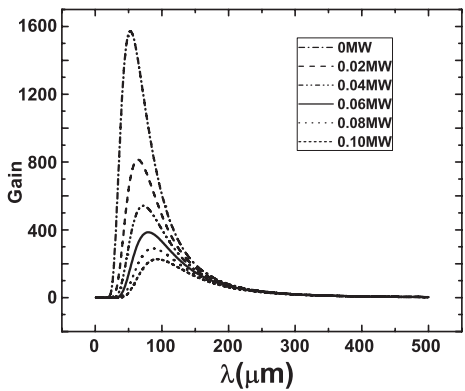}
   \label{subfig:fig3-8}
}
%\subfloat[short for lof][long subfig4 caption]{
%   \includegraphics[width=0.45\linewidth]{figure4.png}
%   \label{subfig:fig4} 
%}
\caption[short for lof]{The gain curves of a noisy and a smoothed beam as functions of modulation wavelength and laser power.}
\label{gaincurve}
\end{figure}

\begin{figure}[htb]
   \centering
   \includegraphics*[width=60mm,angle=0]{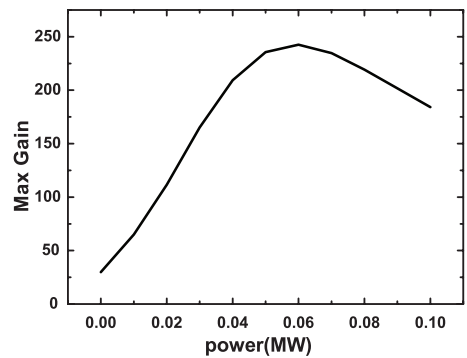}
\vskip-0.10in
   \caption{The behavior of the peak of the gain curve of the noisy beam.}
   \label{gainpeak}
\end{figure}

Currently, a laser heater has been a common device to suppress microbunching instability~\cite{ZHuang} by increasing the uncorrelated energy spread of the electron beam in the linac of a FEL facility. Figure~\ref{laserheater} shows the structure of the SXFEL laser heater including an infrared injection laser of wavelength around 1000 nm, an undulator, and a chicane-type bunch compressor. To study the behavior of the laser heater, the gain curves as functions of laser power are computed separately for the smoothed and noisy beams. Figure~\ref{gaincurve}\subref{subfig:fig1-8} shows the gain curves of a noisy beam as functions of laser power, and Figure~\ref{gaincurve}\subref{subfig:fig3-8} provides those of a smoothed beam as a function of laser power. In comparison of the two plots in Figure~\ref{gaincurve}, we find that as the laser power increases, the peak of the gain curve of the smoothed beam decreases.  This is consistent with the instability theory since the laser increases the initial uncorrelated energy spread of the beam, whereas the gain curve of the noisy beam reaches a maximum at around 0.06 MW before it starts to drop off (Figure~\ref{gainpeak}).

Figure~\ref{gainpeak} can be explained as follows: The cause of the inconsistency comes from the initial noise of the beam current. During beam transportation, the initial current noise is turned into the uncorrelated energy noise of the beam by various impedances such as the LSC impedance, CSR impedance, etc. On the other hand, on top of this, the laser heater introduces extra energy noise, which reduces the gain of the instability after BC1. The result is that the uncorrelated energy spread observed before BC2 goes down simultaneously because of the decrease of the instability gain through BC1.  This explains why the peak of the gain increases as the laser power rises. When the laser heater power becomes higher, the energy noise introduced by the heater dominates and becomes larger and larger. Therefore the slice energy spread before BC2 starts to rise and the magnitude of peak drops with the appearance of a maximum.       

\section{Higher order harmonics of beam current} 

\begin{figure}[htb]
   \centering
   \includegraphics*[width=60mm,angle=0]{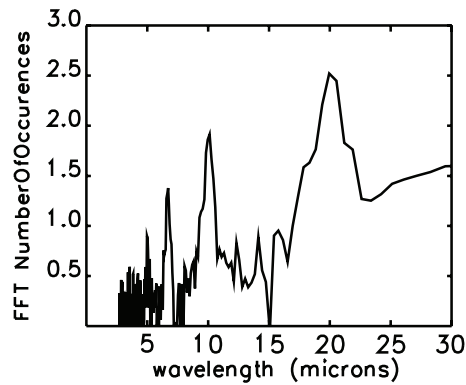}
\vskip-0.10in
   \caption{The fast Fourier transform (FFT) spectrum of the beam current with initial modulation wavelength of 100 $\mu$m at the exit of BC1, where 20 $\mu$m in this figure corresponds to the initial 100 $\mu $m because of the compression ratio of 5.}
   \label{harmonics}
\end{figure}

\begin{figure}[htb]
   \centering
   \includegraphics*[width=60mm,angle=0]{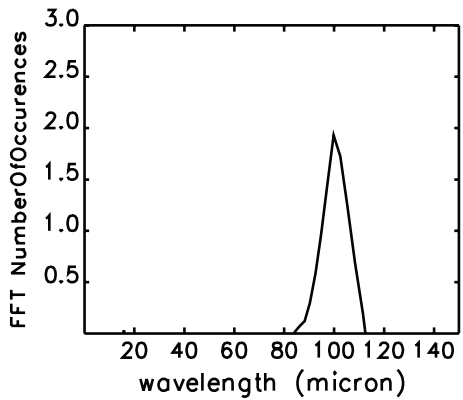}
\vskip-0.10in
   \caption{The fast Fourier transform (FFT) spectrum of the beam current with initial modulation wavelength of 100 $\mu$m at the exit of the injector.}
   \label{initcurFFT}
\end{figure}

Equation~(\ref{saldin}) only takes into account the gain of the fundamental mode of beam current. In other words, the gain curve obtained from it merely describes the behavior of the fundamental mode. However, in our simulations, higher-order harmonics of beam current are also observed along with the initial modulation. We therefore believe that, in certain cases, higher harmonics should be included in the computation of the gain. A rough analysis tell us that the gain of the higher harmonics depends not only on the beam current, but also on the modulation depth. Thus the inclusion of the higher-harmonics becomes a bit more complicated than the fundamental mode. As an example, Figure~\ref{harmonics} shows the excitation of higher harmonics simulated by {\footnotesize ELEGANT} after the first bunch compressor (BC1) with initial modulation wavelength of 100 $\mu$m (corresponding to 20 $\mu$m in Figure~\ref{harmonics} after compression) which is in the vicinity of the peak (Figure~\ref{gaincurve}\subref{subfig:fig1-8}). The initial modulation depth is 5\% in rms and the fast Fourier transform (FFT) spectrum of the initial beam current is illustrated in Figure~\ref{initcurFFT}. We see in Figure~\ref{harmonics} that, besides the peak gain of the fundamental residing around 20 $\mu$m, there are also the first, second, and third harmonics excited at, respectively, 10, 6.7, and 5 $\mu$m. Similar excitations are observed as well at other initial modulation wavelengths such as $80~\mu$m, $60~\mu$m, and $35~\mu$um. Thus we need to keep in mind that higher-order of the initial modulation can also be excited along with the initial modulation. The linear-growth theory should therefore be extended.

\section{Transverse-longitudinal coupling}

One important topic in microbunching-instability study is noise. Like the self-amplified spontaneous emission (SASE) process, micro-bunches can be generated from random noise. If the beam is ideally smoothed in all directions, i.e., without any noise, fluctuation, modulation, etc., microbunching instability can hardly be excited. To reduce microbunching, there are ways to smooth the longitudinal current profile from the cathode by improving the temporal stability of driving laser. However, our investigation indicates that the density noise in the transverse direction can also introduce roughness to the longitudinal beam density distribution as a result of transverse-longitudinal coupling. The reason can be the following: Although it is well-known that the $R_{51}$ and $R_{52}$ elements in the transfer matrix of an ideal chicane-type bunch compressor are both zero, however, these two numbers are not zero for each dipole inside the chicane. As the result, the noise in one transverse plane is transferred into the longitudinal. Although the matrix elements themselves cancel out each other along the chicane, the noise transfer will not be reversed due to its random nature.
\begin{figure}[htb]
   \centering
   \includegraphics*[width=65mm,angle=0]{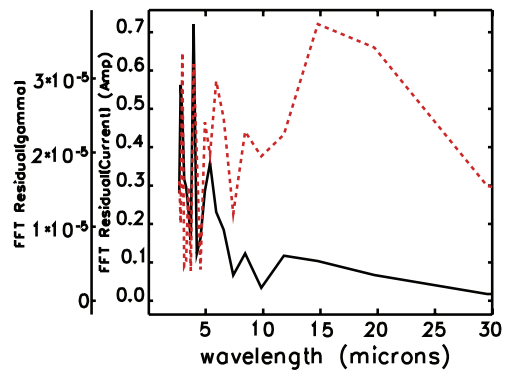}
\vskip-0.10in
   \caption{(Color) The current (black) and energy (red dots) spectra of a beam smoothed in all directions at the exit of BC1 with the laser heater turned off.}
   \label{smoothall}
\end{figure} 

\begin{figure}[htb]
   \centering
   \includegraphics*[width=65mm,angle=0]{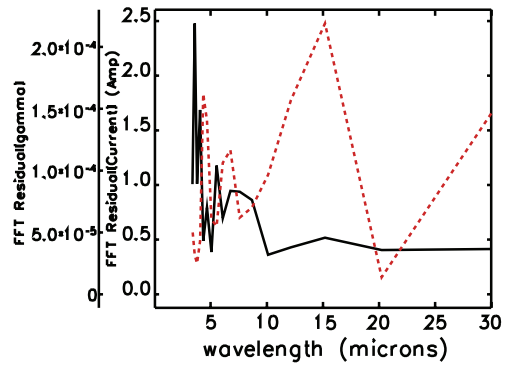}
\vskip-0.10in
   \caption{(Color) The current (black) and energy (red dots) spectra of a beam smoothed only in the longitudinal direction at the exit of BC1 with the laser heater turned off.}
   \label{smoothlong}
\end{figure}

Assuming a beam smoothed in all directions, and another beam with $\thicksim5\%$ noise level in the transverse planes but smoothed longitudinally, Figures~\ref{smoothall} and~\ref{smoothlong} show the current and energy spectra of the beams at the exit of the first chicane (BC1) without laser heating. In the figures, one can see that the levels of density and energy fluctuation (modulation) of the 3-D smoothed beam are undoubtedly smaller than those when the beam is smoothed only longitudinally.   

\begin{figure}[htb]
   \centering
   \includegraphics*[width=65mm]{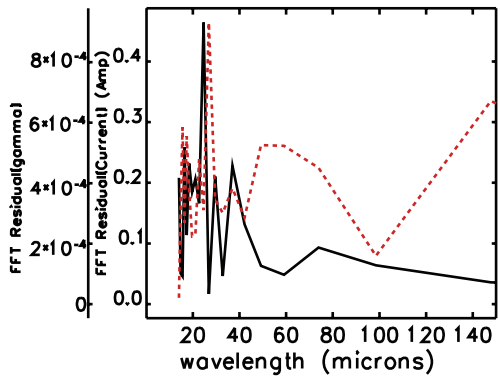}
\vskip-0.10in
   \caption{(Color) The current (black) and energy (red dots) spectra of a beam longitudinally smoothed but transversely noisy at the first BPM right after the laser heater in the SXFEL linac, with the laser heater turned on in the simulation.}
   \label{LH}
\end{figure}

\begin{figure}[htb]
   \centering
   \includegraphics*[width=65mm]{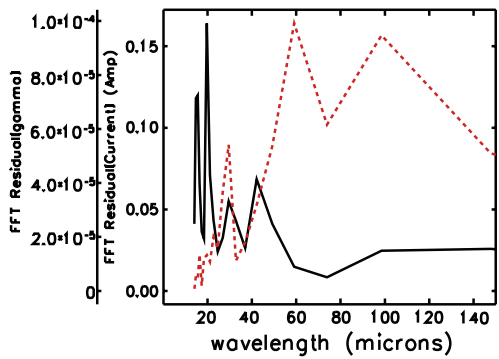}
\vskip-0.10in
   \caption{(Color) The current (black) and energy (red dots) spectra of a beam longitudinally smoothed but transversely noisy at the first BPM right after the location of the laser heater in the SXFEL linac, with the laser heater substituted by a drift in the simulation.}
   \label{noLH}
\end{figure}

Because of the transverse-longitudinal coupling, some extra noise is introduced into the longitudinal direction and the amplitudes of the modulation are also amplified when beam passing through the laser heater. Figures~\ref{LH} and~\ref{noLH} show the current and energy spectra of a beam longitudinally smoothed but transversely noisy at the location right after the designed location of the laser heater with and without the laser heater included in the simulations. We can clearly see that the laser heater introduces extra peaks at various wavelengths and also increases the amplitudes of the current and energy modulation, which will compromise the beam quality thereafter. For this reason, the $R_{51}$ and $R_{52}$ element of the dipoles in the magnetic chicane should be revisited with more care. The parameters of the SXFEL laser heater used in the simulation are: length 0.55 m, periods 10, undulator magnetic peak field 0.31 T, laser peak power 0.80 MW, laser spot size at waist 0.30 mm.

\section{conclusions}
The study of microbunching instability in the SXFEL linac is performed in detail as a classic example. The potential problems and some new effects are uncovered including the initial beam distribution, the higher order modes in the noise amplification, the transverse-longitudinal coupling and the plasma effect in the computation of LSC. More investigations are needed to obtain deeper insights of those problems. 

Computations using both analytic expressions and numerical simulations show that the gain of the microbunching instability indicates large discrepancy between the noisy- and the smoothed-beam input. Since noise in the beam can always be introduced by the random noise, the laser power jitters, etc. in the real case, one should consider how much noise to be included in the initial input. On the other hand, our work shows that in terms of the final density/energy fluctuation (or the bunching factor), the difference between the noisy and the smoothed input is not large.

Higher-order harmonics of the current modulation are also observed in the simulations. This suggests that the linear theory may not be adequate in making estimate of the gain, especially in the vicinity of the peak. The transverse-longitudinal coupling in a magnetic chicane will transport random noise from the transverse dimension to the longitudinal, and this effect needs to be considered in the design of a laser heater. In this sense, we believe that  microbunching instability in an electron linac cannot be completely avoided unless the beam is ideally smooth in the whole 6-dimensional phase space.

Because both the analytic and numerical methods exhibit limitations in the estimate of the microbunching instability, systematic experimental measurements are desired to provide solid and full understanding of the physical essence. Work is ongoing at the Shanghai Institute of Applied Physics (SINAP) to prepare these experiments.

\begin{acknowledgments}
We wish to acknowledge the help of many colleagues in SINAP for the discussions on the analytical and simulation results, and many useful suggestions from the experts in the other institutes. The work is partially supported by National Natural Science Foundation of China (NSFC), grant No. 11275253, and Natural Science Foundation of Shanghai City, grant No. 12ZR1436600. 

\end{acknowledgments}

\end{document}